# Model for molecular absorption spectroscopy in the 1-100 Torr range in the presence of vibrational depletion – Applied to $CH_4$ in $N_2$ and dry air

THOMAS HAUSMANINGER, WEIGUANG MA,[1] OVE AXNER[*]

*Department of Physics, Umeå University, SE-901 87 Umeå, Sweden*
[1]*Present address: State Key Laboratory of Quantum Optics and Quantum Optics Devices, Institute of Laser Spectroscopy, Shanxi University, Taiyuan 030006, China*
[*]*Corresponding author: ove.axner@umu.se*

**When molecules whose collision induced vibrational decay rates are small are probed by molecular absorption spectroscopic (MAS) techniques the absorption signal can, under certain conditions, be reduced and distorted. The reason has been attributed to the fact that a substantial fraction of the molecules in the interaction region will reside in excited vibrational states, which leads to a depletion of the vibrational ground state. One type of molecule in which this can take place is methane. A model for this phenomenon, based on $CH_4$ in trace concentrations in either $N_2$ or dry air in a cylindrical gas cell, detected by mid-infrared light in the 1 – 100 Torr pressure range, is presented. Due to a fast collisional coupling between various rotational states and velocity groups we suggest that depletion in MAS can be modeled adequately by a simple three-level system to which the transport of molecules in the system is coupled as diffusion according to Fick's law, applied to each level individually. The model is verified in a separate work [Hausmaninger T et al., J Quant Spectrosc Radiat Tr. 2017;205:59-70] with good agreement. It predicts that depletion has a strong pressure dependence in the 1 – 30 Torr range, that it is significantly more pronounced in $N_2$ than in air, and that considerable degrees of depletion can be obtained for mW powers of light (> 10% for powers > 20 mW). The findings indicate that, unless precautions are taken, depletion can adversely affect quantitative assessments performed by MAS. Means of how to reduce depletion are given.**

## 1. INTRODUCTION

Laser-based spectroscopic techniques are nowadays frequently used for trace gas detection. Over the years, a number of techniques, with a variety of properties, have been developed for a wide range of applications, not least for trace gas detection. To obtain the highest accuracy, several of these are based upon absorption spectroscopy (AS). Moreover, to achieve the highest detection sensitivity (i.e. to increase the interaction length between the gas to be analyzed and the light), the techniques often incorporate an external cavity [1].

To ensure accurate concentration assessments by such techniques, it is of importance that all phenomena that can influence the measured signal are well understood and that all entities that can influence any such assessment are well known or controlled. If any process that potentially can affect the signal is not fully identified, or if any experimental entity that affects the signal by either a known or a yet-not-identified process is not kept under control (most often kept constant), the technique may suffer from non-linearities or systematic errors, e.g. drifts of signal strengths or invalid calibrations (the latter, e.g., by the use of calibration gas whose composition differs significantly from that of the analyzed gas), that can cause inadequate or even erroneous assessments.

One such example is presented in an accompanying work, in which methane ($CH_4$) at low concentrations in pure nitrogen ($N_2$) was detected in resonant cavities [2]. It was found, for a range of conditions, that the AS signal could be strongly reduced and distorted; in particular, it was found that the reduction and the distortion of the signal increased with pressure up to the 10 Torr range, and that they remained substantial up to at least 100 Torr. It was concluded that these effects cannot straightforwardly be explicated by conventional models for AS. This is clearly an unsatisfactory situation that restricts the use of AS techniques for quantitative assessments; experimental observations like these clearly emphasize the importance of identifying and characterizing all possible phenomena that can affect the AS technique in a not yet fully understood manner so these can be eliminated, circumvented, or compensated for.

When laser-based AS techniques are applied to the detection of molecules, one so far poorly understood and unsatisfactory characterized phenomenon is depletion of molecules in the lower vibrational state addressed by the laser light. This can take place when the light-induced excitation rate constitutes a non-negligible fraction of the vibrational decay rates in the system. As is described in more detail below, in such cases, it is feasible that a substantial fraction of the molecules in the interaction region will reside in excited vibrational states. This can lead to a depletion of the vibrational ground state, manifested as a decrease of the absorption signal.

Adequate models for a similar phenomenon, often referred to as optical saturation (OS) of the laser addressed transition, have been developed for molecular spectroscopy under low pressure (< Torr) conditions [3-6]. However, these models are not valid in all molecular systems and not under many commonly used conditions, in particular not: (*i*) for systems with a "slow vibrational decay", i.e. systems in which the collision induced vibrational decay rates are small (defined below), and (*ii*) for a "short mean free path", i.e. at pressures at which the mean free path of the molecules is shorter than the dimensions of the cavity containing the gas sample. The reason is that these models are built upon a couple of assumptions that are not fulfilled under these conditions, *viz.* that (*1*) all molecules in the laser-illuminated (LI) region that are not in the laser-addressed states or velocity group are in a thermal equilibrium, and (*2*) also all molecules outside the LI region, here referred to as the non-laser-illuminated (NLI) region, are in a state of thermal equilibrium.[1],[2]

The reasons why these two assumptions are not valid in systems in which the conditions (*i*) and (*ii*) are fulfilled are the following.

First of all, in a system with a slow vibrational decay, i.e. when condition (*i*) is valid, the molecules excited by the laser light remain in the excited vibrational state for a considerable amount of time while, at higher pressures, for which also conditions (*ii*) holds, they are redistributed over non-laser addressed rotational states or velocity groups by collisions. This implies that non-laser addressed rotational states or velocity groups cannot be seen as constituting a reservoir in thermal equilibrium, i.e. that assumption (*1*) does not hold. In contrast to the situations when typical OS models are used, which predict a depletion of the lower *rotational* state and the particular *velocity group addressed by the laser light*, conditions (*i*) and (*ii*) can therefore lead to a reduction of the population in the *entire lower vibrational state*, a phenomenon that in this work is referred to as **depletion.** This is similar to the Treanor mechanism that produces significant population in high vibrational states of CO and $H_2$ following vibrational pumping [19].

Furthermore under these conditions it is not necessarily the velocity or rotational state changing collisions that dominate the overall decay rate, as is often assumed in OS models; instead it can be the vibrational deexcitation rates that dominate. Since the latter has a different pressure dependence than the rotational state changing collisions it can be concluded that this phenomenon can bring in a pressure dependence of the absorption signal that many OS models cannot predict.

Secondly, for short mean free paths (*ii*), it can be concluded that also the process that governs the exchange of molecules between the LI and NLI regions is strongly pressure dependent. Under low pressure conditions, for which free flight conditions prevail, all molecules that leave the LI region will propagate in the NLI region until they hit the wall of the gas compartment. After being released from the wall, they will then be in a state of thermal equilibrium, given by the wall temperature. However, at increasing pressures and decreasing mean free paths, the transport process of molecules is slowed down by collisions. For high enough pressures, the transport process can be described as a **diffusion** process for which the transport rate decreases with increasing pressure. This implies that the exchange rate of molecules between the LI and the NLI regions decreases with pressure. For both slow vibrational decay and short mean free paths, i.e. when both the conditions (*i*) and (*ii*) hold, this implies that the molecules are no longer directly transported to the cavity wall and a significant number of molecules in the NLI region around the beam will be in an excited vibrational state. Hence, also assumption (*2*) is not necessarily valid. In this case, it is no longer guaranteed that molecules leaving the LI region are replaced by molecules in a thermal equilibrium.[3]

It is of importance to note that conventional OS models usually assume that there is a fixed (thus pressure independent) transit time for the molecules passing the laser beam, while the diffusion process slows down with increasing pressure. Therefore it can be expected that for the cases when both the conditions (*i*) and (*ii*) are valid, the pressure dependence is also qualitatively different from the prediction of conventional OS models.

As was alluded to above, it has recently been shown, by experiments performed on trace concentrations of methane in nitrogen in the 1 – 100 Torr region, that the absorption signals can be strongly reduced and distorted at pressures in the Torr region [2]. Since this is a gas system for which condition (*i*) holds, and the pressure range used is a range for which condition (*ii*) is valid, traditional OS models cannot be used to explain or predict the observed distortions of the signal;[4] they will predict only an insignificant influence of the laser power on the absorption signal [7, 11]. This is an unsatisfactory situation from several points of view, in particular since methane constitutes a type of molecule whose presence is important to monitor for a multitude of reasons [22, 23] and the pressure range considered is the range that is preferred for highly sensitive trace gas detection [24]. Instead, to be able to predict the influence of depletion of the lowest vibrational state and the pressure dependence of diffusion on the signals measured by molecular AS (MAS) techniques under various conditions, a user-friendly model that incorporates both these phenomena is needed.

---

[1] It is often implicitly assumed in the OS-based models that neither the molecules in the LI region, nor those in the NLI region, are affected by the excitation process; thus that both regions consist of molecules that primarily are in the vibrational ground state

[2] These assumptions (*1*) and (*2*) allow for descriptions of non-linear responses due to OS by the use of simple two level systems. In these cases, the non-laser induced decay rates are often modeled as a sum of molecular collision rates and a transit rate, where the former ones are usually assumed to be given by the rate of rotational state and velocity group changing collisions, while the latter one assumes that every exited molecule is replaced by a molecule in thermal equilibrium with the surrounding at a constant (pressure-independent) rate. For high laser intensities, for which the laser-induced processes start to rival or even supersede the non-laser induced decay rates, these models predict an equilibration of the populations in the two laser addressed rotational states and the velocity group addressed. These models can predict both reduced and distorted Doppler broadened signals [5-12] as well as sub-Doppler responses [1, 5, 7, 11, 13-18].

[3] In fact, the latter process is similar to what happens in some other systems, not least in gas lasers, in which excitations (of atomic or molecular species) can live long enough to counteract lasing. The most well-known situation is the HeNe laser, in which excited Ne atoms have such a long natural lifetime that they need to hit the wall of the gas tube to get deexcited to their ground state. As a consequence, if the gas tube in a HeNe laser is made too wide, the laser will produce less light due to pile up of atoms in the lower laser-connected states [11].

[4] It has been shown that for methane in argon a two level OS description can be applied up to 1.5 Torr [12]. However, since the rotational coupling rate for methane is one order of magnitude stronger in $N_2$ than in Ar, and the vibrational decay rates of methane are slow in both cases [12, 20, 21], it is expected that the OS description is not valid for $CH_4$ in $N_2$ at pressures above 1 Torr.

To be able to model all conceivable phenomena that can take place in a system comprising trace concentrations of a polyatomic molecule in a buffer gas probed by a MAS technique addressing a vibrational-rotational transition, it is generally assumed that the system needs to be described as an multi-level system [25-28]. Since such a system involves a large number of processes and rates, it can be both cumbersome to set up and to use. This often prevents such a model from being used for MAS applied to trace gas analysis. However, for a set of restricting conditions simpler models can be found. Such a model is presented in this work.

The model is mainly restricted to pressure above 1 Torr and trace concentrations (<10 ppm) of the molecule. For this range of pressures there is usually a strong collisional coupling between rotational states and velocity groups within a given vibrational level (or cluster of vibrational levels) [29].[5] Based on this assumption it is argued in this work that it is possible to model depletion in MAS adequately by a simple three-level system in which the laser probing excitation is applied between the lowermost and the uppermost levels. Moreover, because of the short mean free path at these pressures, we assume that the transport process of the molecules in the system can be modeled as diffusion according to Fick's law [30]. Additionally, the model assumes a Gaussian intensity distribution of the laser light with a cylindrical symmetry. The model is therefore formulated as a radial-dependent three-level rate equation system that incorporates diffusion in the radial direction, applied to each level individually. This is a significantly more compact (three-level system), but also more extensive (diffusion in the radial direction, applied to each level individually), way of modeling optical excitations and distribution of population in the molecular system addressed than what a general multi-level description with a fixed transit-time would be able to provide.

Although we use the methane molecule as the pilot molecule in this study and develop the model for this molecule, the type of model presented is not restricted to this species. Based on the considerations presented in this work the model can easily be adapted to other molecular systems adhering to the aforementioned conditions by changing the parameter values used. It can also swiftly be modified, if needed, e.g. by including pressure independent decay rates to take into account significant spontaneous emission or by removing the intermediate level if non-such exist.

In addition to presenting such a model, to illustrate its predictions and in particular show the importance of the vibrational decay rates and the diffusion process and their pressure dependence, this work presents simulations based on the model for various conditions. It includes simulations of the relative reduction of MAS signals due to depletion, which here is referred to as the **assessed degree of depletion.** The latter serves as a measure of the amount of non-linear response of the system due to depletion that can be experimentally assessed and compared with simulations as is done in our accompanying paper [2]. The study illustrates that the model can serve as a basis for not only predicting under which conditions non-linear phenomena associated with depletion and diffusion can appear, it can also be used to identify effective remedies to circumvent the effect of them, e.g. a reduction of laser power, addition of a fast quencher (e.g. $O_2$ or $CO_2$), or a change of the cavity geometry.

The work focuses on methane in $N_2$ and air between 1 and 100 Torr, since it represents the range of pressures that is preferably used for highly sensitive trace gas detection [24]. In addition, experimental evidence of depletion under these conditions have already been demonstrated [2].

To verify the validity of the model presented here, and to experimentally investigate the phenomena predicted, we provided, in our accompanying work, a thorough study regarding detection of $CH_4$ in trace concentrations in an optical cavity by both direct cavity enhanced absorption spectrometry (DCEAS) and noise-immune cavity-enhanced optical heterodyne molecular spectroscopy (NICE-OHMS), addressing its fundamental vibration transitions around 3.4 μm [2]. It is shown in that work that the simple three-level, diffusion-encompassing model presented here can adequately replicate all major features present in the experimental results for the 1 to 100 Torr range [2].

## 2. THEORY

The main aim of this section is to present a simple model that offers an adequate description of MAS signals taking both depletion and diffusion into account. The model should provide information about the dependence of the MAS signals on parameters such as laser power, pressure and type of buffer gas, while still being simple to use. The latter implies that the number of parameters describing the molecular properties should be as low as possible. To achieve this, a few general assumptions and simplifications are made. The validity of these is discussed with respect to the example of methane as this is a well-characterized molecule that is strongly affected by depletion under the conditions considered. While some of the properties of methane are similar to those of other molecules, e.g. a relatively low spontaneous emission rates, other properties, e.g. the existence of three nuclear spin isomers, are specific to this particular molecule. Despite this, the discussion given can serve as a guideline when similar models for MAS under depleted conditions need to be derived for other molecules.

### A. Energy structure of methane

The methane molecule has four vibrational modes: two bending modes, referred to as $\nu_2$ (asymmetric) and $\nu_4$ (symmetric), sometimes jointly termed $\nu_b$, and two stretching modes, denoted $\nu_1$ (symmetric) and $\nu_3$ (asymmetric), together called $\nu_s$. As is shown in Fig. 1, the first excited bending levels, $\nu_4$ and $\nu_2$, which have energies of 1 311 and 1 533 cm$^{-1}$, respectively, have approximately half the energy of the first excited stretching levels, $\nu_1$ and $\nu_3$, which are at 2 917 and 3 019 cm$^{-1}$, respectively. This implies that the vibrational energies of methane form clusters of levels whose energies are separated about 1 500 cm$^{-1}$. The various clusters are referred to as polyads, of which the first one is the dyad, which contains vibrational modes with one quantum of a bending mode, thus either a $\nu_2$ or a $\nu_4$, while the second one is the pentad, comprising both vibration modes with a single quantum of a stretching vibration, thus either a $\nu_1$ or a $\nu_3$, and those with two bending quanta, i.e. $2\nu_2$, $2\nu_4$, or $\nu_2 + \nu_4$, respectively [31-34]. The next polyad, which has an energy of around 4 500 cm$^{-1}$, (not shown in the figure) is denoted the octad and contains levels that comprise either one quantum of a stretching vibration and one of a bending or three bending mode quanta.

---

[5] Strong collisional coupling between rotational states and velocity groups takes place when the collisional redistribution rates are larger than the laser excitation rates for molecules in the laser addressed rotational state and velocity group and implies that the collision redistribution rates within a given vibrational level effectively reduces the phenomenon of "over-" and "under-population" of the rotational states and velocity group addressed by the laser that often take place under low pressure conditions.

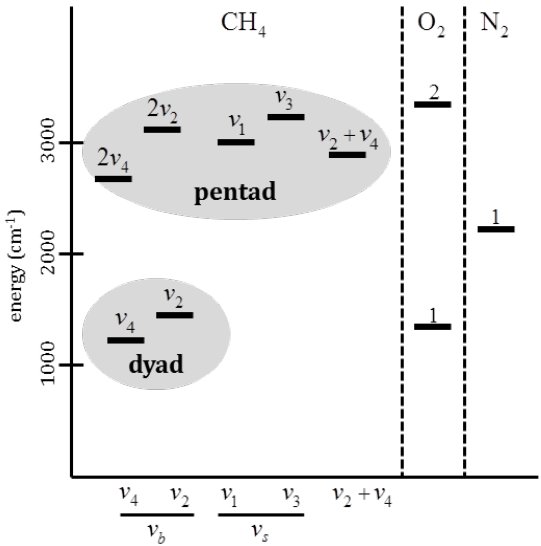

Fig. 1. Vibrational energy level diagrams of $CH_4$, $O_2$, and $N_2$ for energies up to 3500 cm$^{-1}$. The excited energy levels of methane consist of the fundamental bending ($v_2$ and $v_4$) and stretching ($v_1$ and $v_3$) modes and form two clusters of levels referred to as the dyad and the pentad. Not shown in the figure are the symmetry subcomponents and rotational states of the various levels.

Each individual vibration mode comprises a number of rotational states, often denoted by the rotational quantum number $J$ and the symmetry, the latter expressed as $A_1$, $A_2$, $F_1$, $F_2$, or $E$. The symmetries correspond to three nuclear spin isomers with (total) nuclear spin, $\varsigma$, being 2 (for $A_1$ and $A_2$), 1 (for $F_1$ and $F_2$) and 0 (for $E$) [35]. In thermal equilibrium at room temperature the relative abundancies of these isomers, here denoted $\delta_A$, $\delta_F$, and $\delta_E$, are 31%, 56%, and 13%, respectively [36].

The strongest transitions from the ground state to the pentad couple to states with one quantum of $v_3$. This work addresses one of these transitions, from the $0(A_1)J = 7, E, 1$ state in the vibrational ground state to the $1v_3(F_2)J = 6, E, 17$ state in the pentad,[6] at a wavelength of 3.392 μm, which also was chosen as one of the pilot transitions for the experimental study [2].

**B. Means to simplify the energy structure description of methane under collisional dominated conditions**

The energy transfer processes between vibrational and rotational levels are primarily collision induced. Since such processes are energy conserving, they depend on details of how the energy level structure of the molecule under study relates to those of the collision partners. Therefore, for a given species, the energy transfer takes place at dissimilar rates under different environmental conditions.

Population transfer processes in methane have previously been investigated in some detail in a number of studies, revealing a plethora of processes, some of which refer to state-to-state collision rates [20, 21, 29, 32, 36-43]. Although all these studies make it possible to predict the fate of an excited methane molecule under some given conditions, the large number of processes assessed makes such an endeavor extensive and cumbersome. Fortunately, in order to assess the absorption rate of light under a given set of experimental conditions, it is not always necessary to take into account all individual population transfer processes in detail. As is schematically depicted in Fig. 2, it is sufficient in many cases to have knowledge about the magnitude of certain processes since such information can serve as a basis for simplifications of the description of the molecule under study. The aim of the following discussion is therefore to create a simple model, primarily based on the limiting rates in the system, that rather swiftly can describe the excitation and deexcitation processes without too much loss of accuracy.

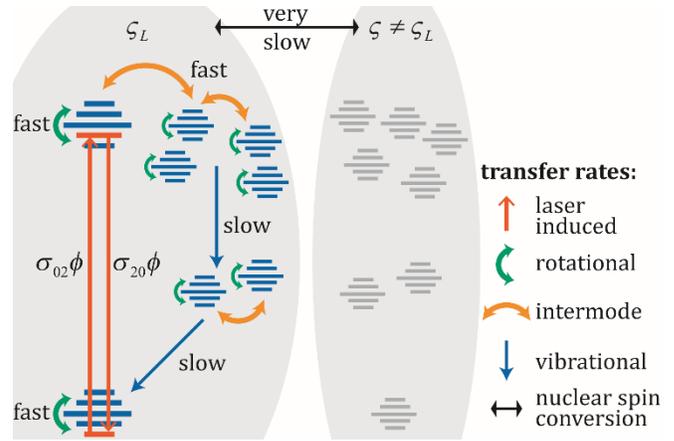

Fig. 2. The model of methane assumes that the nuclear spin conversion rate (black arrow) is slow (whereby only one nuclear spin isomer, $\varsigma_L$, is addressed at a time) while there is fast mixing between different rotational states (green arrows), velocity groups, and vibrational states within a given polyad (orange arrows). The coupling between different polyads (blue arrows) is considered slow and can be in the order of the excitation rate. Entities used are defined in the text.

To predict the fate of excited methane molecules in a cell with gas we will here focus on some specific sets of conditions. As was alluded to above, we will, in particular, consider the case with trace concentrations (i.e. less than a few ppm) of $CH_4$ molecules in a buffer gas, either $N_2$ or dry air, at total pressures of a few Torr up to 100 Torr addressed by a narrow linewidth laser in the 3.4 μm region. The pressure range considered is of particular importance since it has been proven to be suitable for highly sensitive trace gas detection [24].

*1. Coupling of rotational states*

A narrow linewidth laser addresses normally only a single transition, between two specific rotational states, and affects primarily only the populations of the two states addressed. However, it has been found that for $CH_4$ in $N_2$ or $O_2$, at pressures above a few Torr, the collision redistribution rates for a given spin isomer among various rotational levels within a given vibrational state is fast, typically in the order of 10 MHz/Torr [40]. It was therefore concluded by Menard-Bourcin *et al.* [29] that, despite the fact that the laser addresses only a single rotational state, the population of the rotational levels of a spin isomer with spin $\varsigma$ in a given vibrational state can be considered to be in a thermal equilibrium. This implies that the population in a rotational state $j$ of a vibrational state $k$ can be written as

$$N_{k,\varsigma,j} = \frac{g_j}{Q_{k,\varsigma}} e^{-E_j/k_B T} N_{k,\varsigma},\qquad(1)$$

where $g_j$ is the degeneracy of the rotational state, $Q_{k,\varsigma}$ the internal partition sum of the vibration state $k$ of the spin-$\varsigma$ isomer, $E_j$ the rotational energy of the rotational state $j$, $k_B$ the Boltzmann constant, $T$ the temperature, and $N_{k,\varsigma}$ the total number density of molecules with nuclear spin $\varsigma$ in the vibrational state $k$ [44].

---
[6] The assigned rotational quantum numbers represent the angular momentum, the rovibrational symmetry, and a running number according to increasing energies [35].

*2. Coupling of velocity groups*

In addition, when a gas is at sub-atmospheric pressure, any transition addressed at a given frequency by a narrow linewidth laser will, due to the Doppler broadening, be perceived as being (at least in part) inhomogeneously broadened. This implies that light from such a laser will primarily address solely a given velocity group of molecules. Collisions can though also transfer molecules between different velocity groups. This implies that, with a laser beam tuned to a given molecular transition, the molecules will also be transferred in and out of resonance. As above, it can be assumed that for pressures of a few Torr (or above) this velocity-changing collision rate is so fast that it is possible to consider the molecules in any state to be well-distributed also among all possible velocity groups, in this case, according to the ordinary Maxwell-Boltzmann distribution.

*3. Coupling of vibrational states within a polyad*

Moreover, it has been found that also the intermode exchange rates between different vibrational levels within the tetradecade are high, in the order of 1 MHz/Torr [32]. Although not yet specifically assessed for the pentad and the dyad, we will assume that the intermode transfer processes within these polyads are of the same order of magnitude. This implies that it can be assumed that for pressures of a few Torr or above, each molecule will, while being in the interaction region, be exposed to a number of intermode transfer processes. Hence, under these conditions, we will in this work assume that also all vibrational levels within a given polyad are in a thermal equilibrium.

*4. Coupling between the polyads*

On the other hand, it has been concluded by Menard-Bourcin *et al.* that so called vibration-vibration (V-V) energy transfer processes between excited $CH_4$ molecules and $N_2$ molecules are very slow [21]; as is shown in Fig. 1, primarily due to a significant energy mismatch between the energy separations between various vibrational modes in the two types of species. This implies that in an $N_2$ environment, the deexcitation rate from the pentad to the dyad is dominated by vibration-translation/rotation (V-T/R) processes converting one bending quantum to translational or rotational energy (i.e. thermal energy). The same is valid for the dexcitation from the dyad to the ground state. The latter of these has a rate of only ~$10^2$ Hz/Torr; 70 Hz/Torr according to Yardley et al. [20] and 105 ± 15 Hz/Torr according to more recent data of Boursier et al. [21]. By symmetry reasons, the rate between the pentade and dyad is twice as large.

The direct deexcitation rate of the stretching modes in the pentad of $CH_4$ to the ground state is assumed to be much smaller than this as the energy that needs to get converted to translation energy is significantly larger.

In other environments, on the other hand, e.g. in air, which contains also a number of other molecular species, e.g. $O_2$, $CO_2$, and $H_2O$, the decay rates are typically significantly larger as not only V-T/R processes, but also V-V processes contribute to the deexcitation rates. For example, according to Boursier et al., the decay rates from the pentad to the dyad and from the dyad to the ground state are, in a pure oxygen environment, in the kHz/Torr range (9 and 4 kHz/Torr respectively [41]). Although not yet experimentally assessed (to the authors' knowledge), it can be assumed that also the decay rates of excited methane molecules in water vapor are significantly larger than in pure $N_2$, and of similar magnitude as that in $O_2$ [45]

Under trace-gas conditions, i.e. for methane concentrations below 10 ppm, rates for near resonant V-V transfer rates from methane-methane collisions are exceedingly small (<3 Hz/Torr) [41]. This means that it is possible to neglect the influence of such collisions.

Moreover, since the spontaneous decay rates (i.e. the A-factors) for dipole allowed transitions from the pentad are very low (at the most 33 Hz from the pentad and 2 Hz from the dyad [34]), they can be neglected as compared to the collision induced rates at pressures of a few Torr or above. This also implies we can consider the system modeled to be collision dominated.

*5. Nuclear spin conversion*

Yet another process that can affect the system considered is the conversion of molecules between different nuclear spin symmetries. Nuclear spin conversion rates for methane under trace-gas conditions are unfortunately not well characterized. To the authors' knowledge, nuclear spin conversion rates for methane in pure $N_2$ or dry air has not been assessed. On the other hand, it has been found that, for neat methane, they are around 400 Hz at 1 Torr and at most 1 kHz at around 15 Torr [43]. This indicates that it is not directly evident to which extent nuclear spin conversion will affect the population redistribution of methane at various pressures. On the other hand, however, it has been found that the conversion rates for $CH_3F$ is significantly smaller in a buffer gas with $N_2$ or $O_2$ than when neat $CH_3F$ is considered [46]. If the same holds true for methane, the spin conversion rates for methane in $N_2$ and dry air are slower than the V-T rates for collisions with $N_2$ (which are 100 Hz/Torr). This assumption is supported by the fact that experimental data presented in our accompanying work do not show any evidence that nuclear spin conversion plays any significantly role [2]. To facilitate matters, we have here, for simplicity, assumed that nuclear spin conversion can be neglected.

**C. Simplified energy level model of methane for excitations of molecules to a state in the pentad**

*1. Simplified energy level diagram of methane for excitations to a state in the pentad; influence of light-induced processes and collisions*

A comparison of the rates of the various processes described above shows that, under the conditions considered, the states of a given spin isomer in a given polyad (in our case, the pentad, the dyad, and the vibrational ground state) are coupled to each other by rates that are several orders of magnitude larger than the relaxation processes of the vibrational quanta, i.e. those between the various polyads. This implies first of all that it is appropriate to assume that all molecules in a given polyad (and of a given spin isomer) are distributed among all accessible states and velocity group in a thermal equilibrium. This justifies that all levels in a given polyad, corresponding to a given nuclear-spin, can be modeled as a single level. It also implies that the relaxation processes act as a bottleneck, whereby their rates therefore have a strong influence on the populations of various energy levels.

As is illustrated in Fig. 3, the CH$_4$ molecules with a given nuclear spin can therefore be seen as consisting solely of three aggregated levels, henceforth simply termed the "ground state", denoted "0" and representing all states that do not have any vibrational excitation, the "dyad", symbolized by "1" and representing molecules that have one quantum of bending excitation, and the "pentad", denoted "2", representing states with either one quantum of stretching or two quanta of bending.[7]

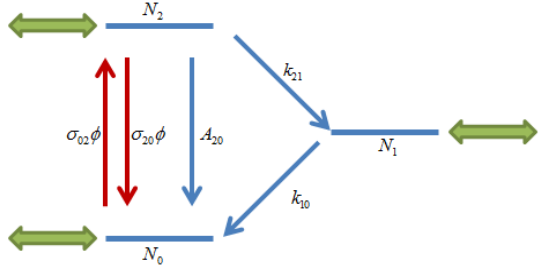

Fig. 3. Simplified energy level diagram for excitation of methane molecules with the nuclear spin addressed by the laser light from a particular state in the vibrational ground state to a specific state in the pentad for pressures of a few Torr and above. The various $N_i$ represent the number densities in the three states of molecules with the particular (total) nuclear spin addressed by the laser.

As was alluded to above, the laser only addresses molecules with a specific nuclear spin, henceforth denoted $\varsigma_L$. Under the condition that $\varsigma_L$ is conserved the model needs only to consider molecules with this particular nuclear spin. Let us therefore, for simplicity, denote the total number density of molecules with nuclear spin $\varsigma_L$ by $N_T$ and the number density of molecules in the three levels of the model that have this particular nuclear spin by $N_i$ (thus for $i=0,1,$ and 2), i.e.

$$N_T = \sum_i N_i = \delta_L N_{CH_4}, \quad (2)$$

where $\delta_L$ is the abundance of the molecules with nuclear spin $\varsigma_L$ while $N_{CH_4}$ is the total number density of CH$_4$ molecules (irrespective of their nuclear spin).

*2. Simplified energy level diagram of methane for excitations to a state in the pentad; influence of diffusion*

As was alluded to above, in addition to the conventional spontaneous, light-induced, and collision-induced processes that can affect the fate of excited molecules also spatial transport of the molecules in the cell is of importance as this can influence the number of excited molecules in the LI region and therefore also the measured absorption signals.

Since the lifetimes of the excited polyads are long, the molecules that have been excited by laser light in the LI region will diffuse into the NLI region and remain in an excited state for a significant amount of time. This implies that the spatial distribution of excited molecules in the sample cell will depend on both the lifetime of the excitation and the processes that transfer molecules between various parts of the cell. Under certain conditions, the excitation is so long lived that the molecules survive in an excited state until they hit the wall of the cell. As any interaction with the wall effectively will lead to deexcitation of the molecules, it is therefore possible that the spatial distribution of the excited molecules in the sample cell will depend also on the geometry (primarily the diameter) of the sample cell.

Since the mean free path of a typical molecule is significantly shorter than the waist of a typical Gaussian beam for gas pressure in the Torr region (it is around 40 µm for N$_2$ at 1 Torr [47], while a typical beam waist is between 0.5 and 1 mm), it is here considered suitable to describe the gas exchange process in the cavity by a diffusion model based on Fick's law [30].

**D. Rate-equations for the level populations**

*1. Rate-equations for the populations of the various levels in the presence of laser induced processes, collisions, and diffusion*

Since both the cavity and the flux distribution of the laser beam have a cylindrical geometry, which are assumed to be co-linear, we will assume that the number densities of molecules in various states of the model have a $\rho$-dependence, where $\rho$ is the radius from the center of the laser beam (or the cavity axis), i.e. that they can be seen as $N_0(\rho)$, $N_1(\rho)$, and $N_2(\rho)$.

Based on the discussion above, and the model of CH$_4$ given by Fig. 3, the rate-equations for the populations of the various states in the model, for a given radial position in the cavity, $\rho$, and at a time $t$, can then be written as

$$\frac{dN_2(\rho,t)}{dt} = \phi(\rho)[\sigma_{02}N_0(\rho,t) - \sigma_{20}N_2(\rho,t)] - k_{21}N_2(\rho,t) + \left[\frac{dN_2(\rho,t)}{dt}\right]_D, \quad (3)$$

$$\frac{dN_1(\rho,t)}{dt} = k_{21}N_2(\rho,t) - k_{10}N_1(\rho,t) + \left[\frac{dN_1(\rho,t)}{dt}\right]_D, \quad (3)$$

and

$$\frac{dN_0(\rho,t)}{dt} = \phi(\rho)[-\sigma_{02}N_0(\rho,t) + \sigma_{20}N_2(\rho,t)] + k_{10}N_1(\rho,t) + \left[\frac{dN_0(\rho,t)}{dt}\right]_D, \quad (4)$$

where $\sigma_{02}$ and $\sigma_{20}$ are the cross sections for absorption and stimulated emission, respectively (both in m$^2$), $\phi(\rho)$ is the flux of the laser light (in number of photons m$^{-2}$ s$^{-1}$), $k_{ij}$ ($i \neq j$) is the collisional induced deexcitation rate from level $i$ to level $j$ (in Hz), while the $[dN_i(\rho,t)/dt]_D$ terms represent the diffusion of the molecules in the state $i$ in and out of the volume element at $\rho$ (in m$^{-3}$ s$^{-1}$). As discussed above, for the pressure conditions addressed, spontaneous emission ($A_{20}$ in Fig. 3) can be neglected.

The cross section for absorption, $\sigma_{02}$ (in m$^2$), defined as in the expressions above, can, in this case, be written as

$$\sigma_{02}(\Delta\nu) = 10^{-4}\frac{S}{\delta_L}\chi(\Delta\nu), \quad (5)$$

---

[7] It was specifically confirmed by numerical simulations based on an extended energy level description of methane that for pressures above 1 Torr and for powers lower than 5 W the fraction of molecules in a specific rotational state and velocity is given according to the partition function, Eq. (1), and the Maxwell-Boltzmann distribution, respectively. This supports our assumption that the fate of excitations in methane, under the aforementioned conditions, can be well described by a three-level system.

where $\Delta\nu$ is the detuning of the laser center frequency from the center frequency of the transition addressed, $S$ is the ordinary (molecular) line strength of the transition [in cm$^{-1}$/(molecule cm$^{-2}$)], and $\chi(\nu)$ the area-normalized absorption line shape function (in cm). The numerical factor of 10$^{-4}$ converts the units of $\sigma_{02}$ from cm$^2$ to m$^2$ while the $\delta_L$ entity takes into account the fact that $S$ is the apparent line strength of the transition addressed when a sample that contains molecules with a distribution of nuclear spins are analyzed, taken from a relevant data base [34], while the absorption cross section solely refers to the ensemble of molecules that have the nuclear spin $\varsigma_L$. The cross section for stimulated emission, $\sigma_{20}(\Delta\nu)$, can, in this case, be related to that of absorption, $\sigma_{02}(\Delta\nu)$, by

$$\sigma_{20}(\Delta\nu) = \frac{Q_0}{Q_2}\sigma_{02}(\Delta\nu), \qquad (6)$$

where $Q_0$ and $Q_2$ are the total internal partition sums of the levels 0 and 2, respectively.

As was alluded to above, for the case when the pressure is sufficiently high (above several Torr), it is common to model the diffusion by Fick's second law [30], which can predict how diffusion causes the concentration to change with space and time and can, for a cylindrical geometry, be written as

$$\left[\frac{dN_i(\rho,t)}{dt}\right]_D = D\frac{1}{\rho}\frac{\partial}{\partial\rho}\left[\rho\frac{\partial N_i(\rho,t)}{\partial\rho}\right], \qquad (6)$$

where $D$ is a diffusion constant (in m$^2$/s) [30].

Although not always fully valid, it is customary to assume that $D$ is inversely proportional to the pressure, i.e. that

$$D = \frac{\tilde{D}}{p}, \qquad (7)$$

where $\tilde{D}$ is a pressure normalized diffusion constant (in Torr m$^2$/s) [48]. We will do the same here.

Under the conditions that the laser beam is collinear with the cavity and that it has a Gaussian flux distribution, $\phi(\rho)$, it can be described as

$$\phi(\rho) = \frac{\Phi_0}{h\nu}\frac{2}{\pi\rho_0^2}e^{-2(\rho/\rho_0)^2} \qquad (7)$$

where $\Phi_0$ is the total power of the beam (in W), $h\nu$ is the photon energy (in J), and $\rho_0$ is a measure of its semi-diameter (in m) [11].[8]

*2. Boundary conditions*

Since all rates considered in the model, under the pertinent conditions, are significantly faster than the typical scanning rates for DCEAS and NICE-OHMS signals, where the latter is in the order of Hz [2], it was assumed that the system will, for each individual wavelength, reach a steady state. This implies that we can solve the system of rate equations above under steady-state conditions, i.e. by assuming that

$$\frac{dN_i(\rho,t)}{dt} = 0 \qquad (8)$$

---

[8] $\rho_0$ is related to the half-width-half-maximum (HWHM) width of the profile, $\Gamma_{HWHM}$, by $\rho_0 = \sqrt{2/\ln 2}\,\Gamma_{HWHM} = 1.7\,\Gamma_{HWHM}$ [11].

for $i = 0, 1$, and 2.

Since the total number density of molecules is assumed to be a constant the set of rate-equations above, Eqs. (3) - (4), can be reduced to a set of only two equations by use of Eq. (2). Moreover, since the diffusion terms comprise second order derivatives, to solve the system, four boundary conditions are needed.

As molecules that are being released from the wall are in a state of thermal equilibrium, with a temperature given by that of the wall, and, since, at room temperature, the thermal energy, $k_BT$, is significantly smaller than the energy needed for vibrational excitation, we can assume that

$$N_1(\rho_{rim}) = N_2(\rho_{rim}) = 0 \qquad (9)$$

while

$$N_0(\rho_{rim}) = N_T, \qquad (10)$$

where $\rho_{rim}$ is the semi-diameter of the gas cell.

Moreover, because of continuity reasons, it can be assumed that the derivative of the number densities of any state in the center of the beam should be zero, i.e. that

$$\frac{\partial N_i(\rho = 0)}{\partial\rho} = 0 \qquad (11)$$

for $i = 0, 1$, and 2.

Since four boundary conditions are needed to solve the set of rate-equations above, it is sufficient to use solely two of the boundary conditions at the rim and two at the center of the beam to solve the rate equations for the number density of molecules as a function of the detuning and the position in the beam, i.e. the $N_0(\Delta\nu,\rho)$, $N_1(\Delta\nu,\rho)$, and $N_2(\Delta\nu,\rho)$.

Furthermore, as $N_T$ is assumed to be a constant, it is possible to normalize the populations in the expressions above by $N_T$. Practically, this implies that the absolute populations can be replaced by normalized population, $\bar{N}_i$, defined as $N_i/N_T$, for which, according to Eq. (2)

$$\sum_{i=0,1,2}\bar{N}_i = 1. \qquad (12)$$

**E. The absorption signal, the absorption coefficient, and the assessed degree of depletion**

From the solution of the rate equations given above, i.e. based on the $N_0(\Delta\nu,\rho)$ and $N_2(\Delta\nu,\rho)$, it is possible to estimate the change in flux of the light for a given laser detuning at a given position in the beam, denoted $\Delta\phi(\Delta\nu,\rho)$, as

$$\Delta\phi(\Delta\nu,\rho) = \phi(\rho)\Big[\sigma_{02}(\Delta\nu)\bar{N}_0(\Delta\nu,\rho) \\ -\sigma_{20}(\Delta\nu)\bar{N}_2(\Delta\nu,\rho)\Big]N_TL, \qquad (13)$$

where $L$ is the length of the interaction region (in m).

The entity measured in absorption spectrometry is the change in power of the entire beam, $\Delta\Phi(\Delta\nu)$, which is given by

$$\Delta\Phi(\Delta\nu) = 2\pi h\nu \int_0^{\rho_{rim}} \Delta\phi(\Delta\nu,\rho)\rho d\rho. \quad (13)$$

In the case of no depletion (for which we can assume $\bar{N}_0 = 1$ and $\bar{N}_2 = 0$), $\Delta\Phi^{ND}(\Delta\nu)$ is simply given by

$$\Delta\Phi^{ND}(\Delta\nu) = 2\pi h\nu \int_0^{\rho_{rim}} \phi(\rho)\sigma_{02}(\Delta\nu)N_T L\rho d\rho \quad (14)$$
$$= \Phi_0 \sigma_{02}(\Delta\nu) L \delta_L N_{CH_4},$$

where we in the last step have assumed that $\rho_{rim} \gg \rho_0$, used the normalization condition of $\phi(\rho)$, and the fact that $N_T = \delta_L N_{CH_4}$.

To assess the influence of depletion on the detected signal, it is convenient to define the assessed degree of depletion, $\gamma(\Delta\nu)$, as

$$\gamma(\Delta\nu) \equiv \frac{\alpha^{ND}(\Delta\nu) - \alpha(\Delta\nu)}{\alpha^{ND}(\Delta\nu)} = \frac{\Delta\Phi^{ND}(\Delta\nu) - \Delta\Phi(\Delta\nu)}{\Delta\Phi^{ND}(\Delta\nu)}, \quad (15)$$

where $\alpha(\nu)$ is the absorption coefficient under the pertinent conditions, which according to Beer's law, is given by (in m$^{-1}$)

$$\alpha(\Delta\nu) = -\frac{1}{L}\ln\left(1 - \frac{\Delta\Phi(\Delta\nu)}{\Phi_0}\right) = \frac{1}{L}\frac{\Delta\Phi(\Delta\nu)}{\Phi_0}, \quad (16)$$

where the last steps is valid for small degrees of absorption, and where $\alpha^{ND}(\Delta\nu)$ is the corresponding entity in the non-depleted case, which, under the same conditions, can be written as $(1/L)\Delta\Phi^{ND}(\Delta\nu)/\Phi_0$. This implies that $\gamma(\Delta\nu)$ represents the fraction by which the non-depleted absorption is reduced by depletion, which, for the case with small degrees of absorption, is given by the last step in Eq. (15). This also implies that the depleted absorption coefficient can be written in terms of the assessed degree of depletion as

$$\alpha(\Delta\nu) = [1 - \gamma(\Delta\nu)]\alpha^{ND}(\Delta\nu). \quad (17)$$

It is here suitable to note that while the various populations refer to specific positons in the sample compartment for a given laser frequency, i.e. $N_i(\Delta\nu,\rho)$, the change in power of the entire beam, $\Delta\Phi(\Delta\nu)$, the absorption coefficient, $\alpha(\nu)$, and the assessed degree of depletion, $\gamma(\Delta\nu)$, are entities that refer to the entire beam. They can therefore be seen as entities that are averaged over the entire LI region, i.e. basically being weighted by the flux profile of the laser. This can be further illustrated by rewriting $\gamma(\Delta\nu)$ by the use of the definitions above as the first step of

$$\gamma(\Delta\nu) = 1 - \frac{4}{\rho_0^2}\int_0^{\rho_{rim}}\left[\bar{N}_0(\Delta\nu,\rho) - \frac{Q_0}{Q_2}\bar{N}_2(\Delta\nu,\rho)\right]e^{-2(\rho/\rho_0)^2}\rho d\rho$$
$$= \frac{4}{\rho_0^2}\int_0^{\rho_{rim}}\left[1 - \bar{N}_0(\Delta\nu,\rho)\right]e^{-2(\rho/\rho_0)^2}\rho d\rho,$$
(18)

where the $\bar{N}_i(\Delta\nu,\rho)$ are given by the solution to the rate equations.

For limited amounts of depletion, for which $\bar{N}_2 < \bar{N}_0$, and in particular when the partition sum of the upper laser-connected vibrational state is markedly larger than that of the ground vibrational state, the term representing the stimulated emission plays a relatively small role in the system. Under the condition that this holds, and that $\rho^{rim} \gg \rho^0$, the expression for $\gamma(\Delta\nu)$ can be simplified according to the last step in Eq. (18). This shows that, under these conditions, it is possible to interpret the assessed degree of depletion as the reduction of the population of the vibrational ground state averaged over the LI region.

## 3. SIMULATION PROCEDURES

When MAS is used for quantitative assessments, it is of importance to assess under which conditions depletion starts to affect the measurements. To assess the degree of depletion, including the influence of collisions and diffusion on this, it is suitable, based on the model above, to simulate the number density of molecules in the three levels considered at various radii from the center of the axis of the laser beam. Such simulations were therefore performed, based on the rate equations, Eqs. (3) - (4), with the diffusion terms given by the Eqs. (6) and (7), solved under steady-state conditions. The rate equations were solved numerically by a MATLAB finite difference code that implements the three-stage Lobatto IIIa formula (using the MATLAB `bvp4c` function). The simulated populations were then used to calculate $\gamma(\Delta\nu)$ by Eq. (18) under various conditions.

The collision decay rate between the states $i$ and $j$, $k_{ij}$, was calculated as

$$k_{ij} = \sum_l \tilde{k}_{ij}^l \cdot p_l = p_T \sum_l \tilde{k}_{ij}^l \cdot c_l, \quad (18)$$

where $\tilde{k}_{ij}^l$ is the pressure normalized collision rate constant with collision partner $l$, $p_l$ the partial pressure of $l$, $p_T$ the total pressure, and $c_l$ the concentration of species $l$. For dry air, a mixture of 80% nitrogen and 20% oxygen was assumed. All pressures given below refer to the total pressure of the gas mixture.

The line shape function was considered to be a Voigt function (with a pressure broadening coefficient, $\Gamma_p$), since this reasonably well can account for both the effects of Doppler and pressure broadening [11]. The numerical entities used for the simulations presented below are summarized in Table 1.

Since the model is predominantly valid when the mean free path of the molecules is significantly smaller than the beam diameter, and the rotational state changing collision rates are high, simulation were made for pressures for 1 Torr and above. As the transition between the diffusion- and the collision-dominated regime of depletion takes place predominantly for a pressure below 50 Torr (see below), to cover both regimes, simulations were performed for pressures up to 100 Torr.

**Table 1. Numerical values of the entities used in the simulations.**

| Parameter | Value | Unit | Reference |
|---|---|---|---|
| $\tilde{k}_{21}^{N_2}$ | 200 | Hz/Torr | [20, 21] |
| $\tilde{k}_{10}^{N_2}$ | 100 | Hz/Torr | [20, 21] |
| $\tilde{k}_{21}^{O_2}$ | 9 × 10$^3$ | Hz/Torr | [41] |
| $\tilde{k}_{10}^{O_2}$ | 4 × 10$^3$ | Hz/Torr | [41] |
| $\Gamma_p$ | 3 × 10$^6$ | Hz/Torr | [34] |
| $S$ | 3.4 × 10$^{-20}$ | cm$^{-1}$/(molecule cm$^{-2}$) | [34] |
| $Q_0 / Q_2$ | 1/4 | | [34] |
| $\delta_L$ | 0.13 | | [36] |
| $\tilde{D}$ | 0.018 | Torr m$^2$/s | [48] |
| $\rho_0$ | 0.7 × 10$^{-3}$ | m | [2] |

The model presented here was validated in our accompanying paper in which four different transitions in all three spin isomers of methane and two isotopologues ($^{12}CH_4$ and $^{13}CH_4$) were addressed experimentally. It was found that the agreement between the experimentally found and theoretically predicted degrees of depletion was good and that all major features of the signals could be reproduced well, both qualitatively and quantitatively. Because of this, we found it sufficient in this work to scrutinize the phenomena that can take place in the system modelled by addressing a single transition (between the $0(A_1)J=7,E,1$ and the $1\nu_3(F_2)J=6,E,17$ states). Hence, all simulations in this work addresses molecules with a given rotational symmetry, $E$, which has the nuclear spin abundance ($\delta_E$) of 13 % [36].

Moreover, since the experiments performed in our accompanying paper [2] are made by the use of a cavity enhanced technique, intracavity powers up to the Watt region could be obtained by the use of low power (mW) laser systems. Because of this, the powers investigated in this work range from low mW to hundred Watts. However, since the description of the depletion of methane molecules given here is valid irrespective of whether or not an external cavity is used, and since such powers also can be obtained in direct absorption spectrometry, this study will be relevant to also such experimental conditions.

## 4. RESULTS

**A. Level populations**

*1. Influence of collisions and diffusion on the radial distribution of the number density of molecules in the laser-addressed vibrational states*

The model includes two pressure dependent processes that affect molecules in the LI region, and thereby the absorption signal, *viz.* deexcitation by collisions with the buffer gas and diffusion in and out of the LI region, where, in the latter case, the molecules are allowed to diffuse all the way to the cavity wall where they physically lose their excitation energy. To illustrate the effects of these processes on the state populations of methane, Fig. 4 shows a simulation of the radial distribution of the normalized populations of the pentad and the vibrational ground state of the laser addressed nuclear spin isomer, $\bar{N}_2$ and $\bar{N}_0$, for three model systems considering different combinations of processes. The orange curves represent the normalized populations in the absence of diffusion (i.e. for $D=0$), i.e. when solely deexcitation by collisions (with $N_2$) is taken into account, the green curves indicate the situation in the absence of deexciting collision (i.e. for $k_{ij}=0$), i.e. when the deexcitation of the molecules, transferred by diffusion, takes place only by collisions with the cavity wall, while the black curves, finally, illustrate the situation when the molecules are exposed to both collisions and diffusion. The solid curves represent $\bar{N}_0$ while the dashed curves show $\bar{N}_2$ as function of $\rho$ (all left axis). The dotted gray curve (right axis) represents the flux distribution of the laser beam. The molecules were considered to be exposed to a laser beam with a flux distribution given by Eq. (7), with $\rho_0$ and $\Phi$ being 0.7 mm and 30 mW, respectively. The total pressure was assumed to be 15 Torr.

A comparison of these model systems shows that under the conditions considered the state populations of the molecules are affected by both collision and diffusion processes, although to different degrees at different positions.

For the system considering solely collisions, which is illustrated by the orange curves, virtually all molecules outside the LI region (i.e. for $\rho \geq 0.7$ mm) reside in the lowest vibrational state. The reason for this is that there is no transfer of molecules in and out of the excitation region. The fact that the depletion "hole" has a shape that differs from that of the laser beam is due to the fact that the excitation rate is so high that the population of the excited vibrational states is optically saturated.

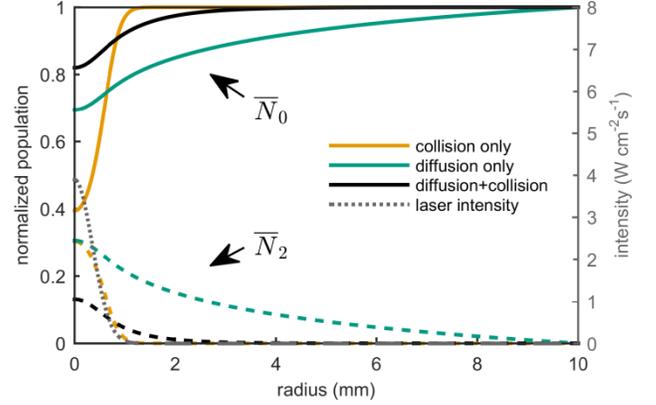

Fig. 4. Simulation of the radial distribution of the normalized populations of the vibrational ground state, $\bar{N}_0$, (solid curves) and the upper laser-connected state, $\bar{N}_2$, (dashed curves) of methane molecules with nuclear spin 0 (left axis) exposed to a laser beam with an photon flux distribution given by Eq. (7), with $\rho_0$ being 0.7 mm and $\Phi_0$ 30 mW, at 15 Torr, for three model systems with different processes taken into account, i.e. deexcitation by collisional deexcitation with $N_2$ (orange curves), diffusion followed by deexcitation at the cavity wall (green curves), and both together (black curves). The flux distribution of the laser beam is indicated by the dotted gray curve (right axis).

For the case when the deexcitation of molecules in the LI region is solely mediated by diffusion, shown by the green curves, there are excited molecules in the entire sample compartment, i.e. also far outside the LI region. The reason for this is that, in this case, the excited molecules can only be deexcited through collisions with the cavity walls.

When both processes are present, finally, displayed by the black curves, excited molecules still exist outside the LI region, although not necessarily all the way out to the cavity wall. The reason for this is that they, to a certain extent, can become deexcited by collisions before they have diffused to the wall. It is obvious that under these conditions, both collisions and diffusion have a significant impact on the populations in the LI region. This implies that both processes have a direct influence on the absorption of light and thereby the detected signal. It also shows that it is not appropriate, under the pertinent conditions, to model the transport processes of molecules in and out of the LI region simply as a constant rate, as often is done in OS models.

*2. Influence of pressure on the radial distribution of the number density of molecules in the laser-addressed vibrational states*

Since the two types of deexcitation processes depend differently on the pressure, the populations of the various vibrational states do not only depend on the laser beam parameters, e.g. the position in the beam and its flux, they also depend on pressure. Fig. 5 shows simulations of the radial distribution of the normalized populations in the two laser addressed states, i.e. $\bar{N}_0$ (solid curves) and $\bar{N}_2$ (dashed curves), for three different pressures, 5, 15, and 80 Torr, represented by the green, the black, and the orange curves, respectively (all left axis). The gray dotted curve represents the flux distribution of the laser beam (right axis).

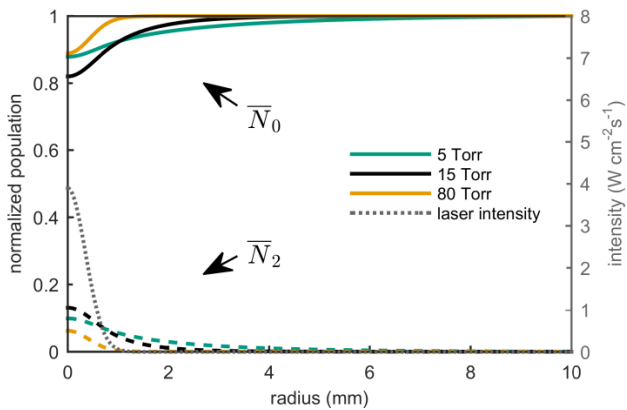

Fig. 5. Radial distribution of the normalized populations of the vibrational ground state, $\bar{N}_0(\rho)$, (solid curves) and the upper laser-connected state, $\bar{N}_2(\rho)$, (dashed curves) of methane molecules (left axis) exposed to a Gaussian laser beam with a power of 30 mW for three different total pressures, *viz.* 5 Torr (green), 15 Torr (black), and 80 Torr (orange). The dotted, gray curve represents the intensity profile of the laser beam (right axis).

For the lowest pressures (5 Torr, green curves) the radial distribution of the population shows a strong similarity to the distributions modeled considering diffusion only (green curves in Fig. 4). The vibrational ground state, $\bar{N}_0$, (green solid curve) is significantly depleted in the LI region (~10 %), and approaches gradually thermal equilibrium ($\bar{N}_0 = 1$) at the cavity wall. The reason for this is that at low pressures the molecules can move freely and the collision rates are low. This implies that under low pressure conditions, the populations of the various levels depend on the geometry of the sample cell.

As the pressure increases, however, (black and orange curves) the populations of the molecules in the outer part of the cell approach thermal equilibrium; for example, for a pressure of 15 Torr (black curves), 99% of the molecules are in the lower vibrational state at a distance of 3 mm from the center of the laser beam. For a pressure of 80 Torr (orange curves), the corresponding distance is 1 mm. For the latter case, the overall radial distribution is close to the one resulting from the system modeled considering collisions only (orange curve Fig. 4). The reason for this is that the diffusion processes become slower and the deexciting collision rates increase with increasing pressure, which implies that the molecules have a successively higher probability of being deexcited by collisions before they have diffused any appreciable distance into the NLI region. As long as the gas cell has a semi diameter larger than these values, the populations of the various levels are practically independent of the geometry of the sample cell.

### B. The assessed degree of depletion and the absorption rate

What matters for absorption spectroscopy is to which extent the alterations of the populations of the various levels affect the absorption signal. As was alluded to above, a measure of this is the assessed degree of depletion, $\gamma(\Delta\nu)$. It is therefore of interest to investigate this entity in more detail.

#### 1. Influence of collisions and diffusion on the pressure dependence of the assessed degree of depletion

Since the two deexcitation processes considered affect the number density of molecules in the various levels differently under different conditions, it is suitable to first assess their separate effects on the assessed degree of depletion. Fig. 6 displays therefore the pressure dependence of the assessed degree of depletion on resonance $\gamma(\Delta\nu=0)$, as given by Eq. (15), for a specific case (a light field with a power of 500 mW), for the same three cases as were displayed in Fig. 4, i.e. for $CH_4$ in neat $N_2$ when the deexcitation of the molecules is solely caused by collisions (orange dashed curve), when it only takes place by diffusion followed by collisions with the cavity wall (green dash-dotted curve), and when both collisions and diffusion act jointly (black solid curve).

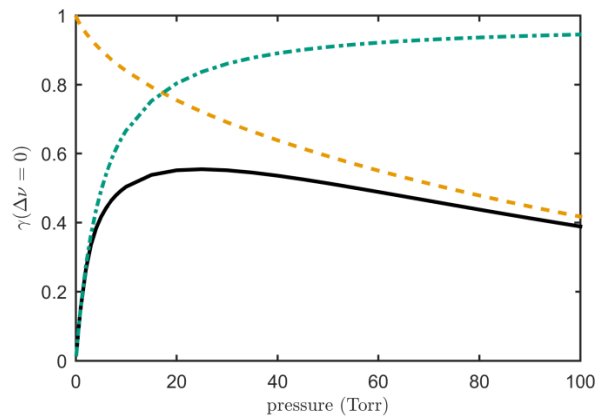

Fig. 6. Pressure dependence of the on-resonance assessed degree of depletion, i.e. $\gamma(\Delta\nu=0)$, for methane in neat $N_2$. The curves show the results of simulations including deexcitation solely by collisional deexcitation (orange, dashed curve), only by diffusion followed by deexcitation at the wall (green, dash-dotted curve), and both (black curve). The power of the light was set to 500 mW for all simulations.

The figure shows that when deexcitation solely takes place by collisions (the orange curve) the assessed degree of depletion is huge for low pressures, while it thereafter *decreases* with pressure. The reason for this is that for low pressures, the deexcitation rates are small, whereby the molecules get trapped in the upper vibrational state. As the pressure increases, the deexcitation rates increase. This implies, according to Eq. (17), that the absorption coefficient, and thereby the measured absorption signal, can be very small. When deexcitation takes place only by diffusion (the green curve), on the other hand, the assessed degree of depletion *increases* with pressure. This is due to the fact that the effective diffusion rate of the excited molecules out of the laser beam decreases the higher the pressure.

It can be assumed that the deexcitation mechanism that, for each pressure, dominates limits the assessed degree of depletion at that particular pressure. From a comparison between the solid curve and the two dashed ones in Fig. 6, it is therefore possible to conclude that, for low pressures (under the conditions considered, for a few Torr) the deexcitation is mainly given by the diffusion process whereby the molecules are primarily deexcited by collisions with the cavity wall. This implies that for the lowest pressures, it is possible to simplify the system by neglecting the collision terms.

At higher pressures, on the other hand (in this case, predominantly above some tens of Torr), the deexcitation is dominated by collisions with surrounding molecules (the diffusion processes play in this case a minor role). This implies that the model can be radically simplified by neglecting the diffusion terms in the system of rate equations, whereby it can be described as a system solely incorporating collisional processes. This also makes it possible to analytically solve the equations either for every $\rho$ position independently (i.e. for the particular laser flux at each $\rho$ position) or for an average laser flux.

In between these pressures, and particularly around 20 Torr, both processes influence the assessed degree of depletion, whereby none can be neglected.

*2 Pressure dependence of the assessed degree of depletion under various power and buffer gas conditions*

The pressure dependence of the on-resonance assessed degree of depletion, $\gamma(\Delta\nu=0)$, in the presence of both collisional deexcitation and diffusion, was then simulated for three different powers of the light and for two types of buffer gases. The results of the simulations are presented in Fig. 7. While the solid curves refer to a buffer gas of pure $N_2$, the dashed ones correspond to dry air. The three different colors, black, green, and orange, correspond to light powers of 30, 100, and 500 mW, respectively.

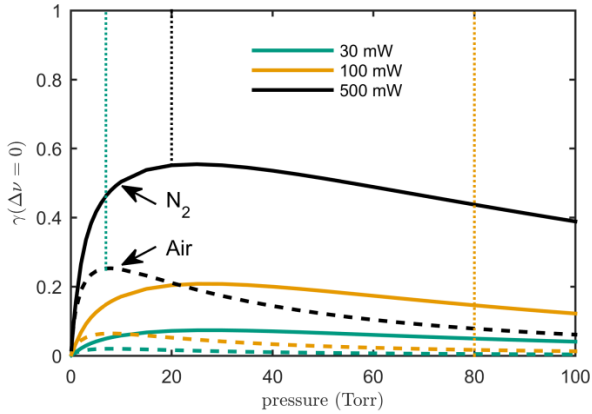

Fig. 7. Assessed degree of depletion of methane as a function of pressure for pure $N_2$ (solid curves) and dry air (a mixture of 80% $N_2$ and 20% $O_2$, dashed curves) as buffer gases for powers of the light of 30 (green), 100 (orange), and 500 mW (black).

The data show first of all that the assessed degree of depletion depends strongly on the power of the light; it is significantly larger for a power of 500 mW (the black curves) than for 100 mW (the orange curves), which in turn is markedly larger than for 30 mW (the green curves). The curves also illustrate, in accordance with the data in Fig. 6, that the assessed degree of depletion has a pronounced pressure dependence; it increases with pressure for low pressures until it reaches a maximum after which it gradually decreases. This implies that depletion will affect the absorption signal significantly for virtually all pressures except the lowest ones.

More specifically, Fig. 7 shows (again for the conditions considered) that while the assessed degree of depletion increases with pressure up to some tens of Torr when the buffer gas consists of neat $N_2$ (the solid curves), it does so only up to around a few Torr when the buffer gas is dry air (the dashed curves). For the particular case with a power of 500 mW, as is shown by the black curves, for the cases with methane in neat $N_2$ and in dry air, illustrated by the solid and dotted curves, respectively, the assessed degree of depletion reaches its maximum for pressures of 20 and 7 Torr, respectively. For pressures above the ones that provide the maximum depletion, the assessed degree of depletion decreases as a function of pressure, although often solely weakly. It also shows that the assessed degree of depletion can be significant, above 50% in neat $N_2$, for a range of pressures. This implies, according to Eq. (17), that the depletion phenomenon reduces the absorption signal to less than 50% of its undepleted value.

*3. Power dependence of the assessed degree of depletion under various pressure and buffer gas conditions*

To estimate at which light power depletion starts to affect absorption signals, the on-resonance assessed degree of depletion, $\gamma(\Delta\nu=0)$, was simulated as a function of power for three different pressures, *viz.* 7, 20, and 80 Torr, shown by the green, black, and orange curves in Fig. 8, respectively. Again, the solid curves represent simulations for neat $N_2$ while the dashed ones correspond to simulations with dry air as buffer gas. The two lowermost pressures, i.e. 7 and 20 Torr, correspond to the pressures for which the depletion is maximum for methane in $N_2$ and dry air, respectively, as is indicated by the vertical dotted lines in Fig. 7, while the highest one, i.e. 80 Torr, is well into the collision dominated regime (as shown by Fig. 6).

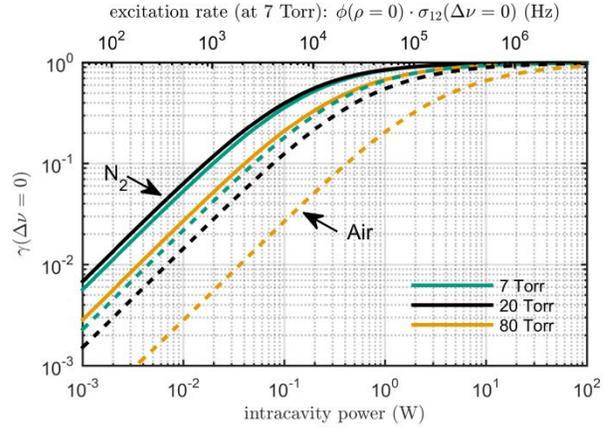

Fig. 8. Assessed on-resonance degree of depletion of methane, i.e. $\gamma(\Delta\nu=0)$, as a function of power of the light (lower axis) and the corresponding on-resonance excitation rate in the center of the beam for 7 Torr (upper axis) for two different buffer gases, pure $N_2$ (solid curves) and dry air (dashed curves), for three different pressures, 5 Torr (green), 15 Torr (black), and 80 Torr (orange).

The log-log plot in Fig. 8 shows that, for all conditions considered, the assessed degree of depletion is increasing linearly with the power of the light, i.e. as $\gamma(\Delta\nu=0) = c_p^{N_2/air}\Phi_0$, up to an assessed degree of depletion of around 20%. However, since the proportionality constant, i.e. the $c_p^{N_2/air}$, which thus is a measure of the degree of depletion for a given laser power, includes the combined effect of collisions and diffusion, depends on both the pressure and the type of buffer gas used ($N_2$ or dry air), it differs significantly between different conditions.

The figure shows that for 80 Torr the ratio between the degrees of depletion when methane resides in neat $N_2$ and when it is in dry air is about one order of magnitude (i.e. $c_{80\,Torr}^{N_2} / c_{80\,Torr}^{air} \approx 10$). For lower pressures they differ by less; for pressures of 20 and 7 Torr by a factor of 3 and 2, respectively (i.e. $c_{20\,Torr}^{N_2} / c_{20\,Torr}^{air}$ and $c_{7\,Torr}^{N_2} / c_{7\,Torr}^{air}$ are 3 and 2, respectively). These differences can be explained by the fact that for high pressures the depletion is mainly given by the collisional de-excitation rate, which is about a factor of 10 larger for dry air than neat $N_2$, while for low pressures it is given to a large part by diffusion, which is assumed to be same for both types of buffer gas. Thus, the lower the pressure, the more does the $c_p^{N_2} / c_p^{air}$ ratio approach unity.

The fact that the assessed degree of depletion depends on the power of the light implies, according to Eq. (17), that also the absorption coefficient, and thereby the relative absorption, change with power of the light. However, for small amounts of depletion the change is weak. For example, for an assessed degree of depletion of 5%, a power variation of 5 % results in a change of the relative absorption of only 0.25%. This can in many cases be neglected.

Also the fact that the assessed degree of depletion depends also on the composition of the buffer gas implies that the absorption signal can change if the composition of the buffer gas changes. However, also this dependency is relatively weak for small amounts of depletion. For example, if the oxygen content in dry air changes from 20% to 18%, for a power of 40 mW and for a pressure of 80 Torr (for which $\gamma = 1\%$ ), the absorption signal is expected to vary less than 0.1%. For the case when $\gamma = 10\%$ the absorption signal variation is less than 1% for the same change of oxygen concentration.

Similarly, it can be concluded that although the deexcitation rates of methane in water vapor is not known, for the case with methane in humid air, and for the case with a relative humidity of 100% at 20 °C, at which the concentration of water in air is 1.4%, if the collision rate constant for methane with water would be the same as that for oxygen, and under the same experimental conditions (i.e. a power of 40 mW and for a pressure of 80 Torr), the absorption signal is expected to vary from that for dry air by less than 0.1%, and less than this for air with lower relative humidity. For lower pressures, for which the deexcitation is dominated by diffusion, a corresponding change in oxygen or water content has no significant influence on the absorption signal. This suggests that the influence of humidity on the absorption signal due to the depletion phenomenon is weak.

On the other hand, if the calibration is based on a reference gas with neat $N_2$ when methane in atmospheric samples is measured, the error due to depletion can be larger, around 10% for the case with 40 mW of light and measurement performed in dry air at 80 Torr, and even larger for other conditions.

*4. Absorption profiles in the presence and absence of depletion*

As is indicated by the model given above, the absorption rate, and thereby the assessed degree of depletion, $\gamma(\Delta \nu)$, do not only depend on the laser power, but also on the detuning of the laser light from the center of the transition, $\Delta \nu$. Fig. 9 shows, in panel (a), a simulation of the assessed degree of depletion for the pilot transition at a pressure of 10 Torr and a power of 3 W as a function of detuning. The figure shows that the assessed degree of depletion indeed has a strong frequency dependence and that its shape is not that of a conventional line shape function.

Panel (b) displays, by the green curve, an area normalized Voigt absorption line shape function for the same general conditions (representing the absorption profile in the absence of depletion). The orange curve in the same panel illustrates the corresponding profile for the case when depletion is accounted for according to the model given above. As can be seen, the two curves differ markedly. This implies that in the presence of depletion the absorption line shapes will differ from the conventional ones.

The line profile will also differ considerably from that when the transition is affected solely by optical saturation. To illustrate this, a Voigt profile, calculated under the condition of strong saturation (assuming a dependence on the degree of saturation, $G$, of $(1+G)^{-1/2}$ and a $G$ of 7.5), which can appear under low pressure conditions, is shown by the black dashed curve. This shows that in the presence of depletion, the absorption profile, shown by the orange curve in panel (b), does not only become significantly smaller than the area normalized Voigt absorption profile that represents the case with no depletion, it has also a shape that differs significantly from that of OS.

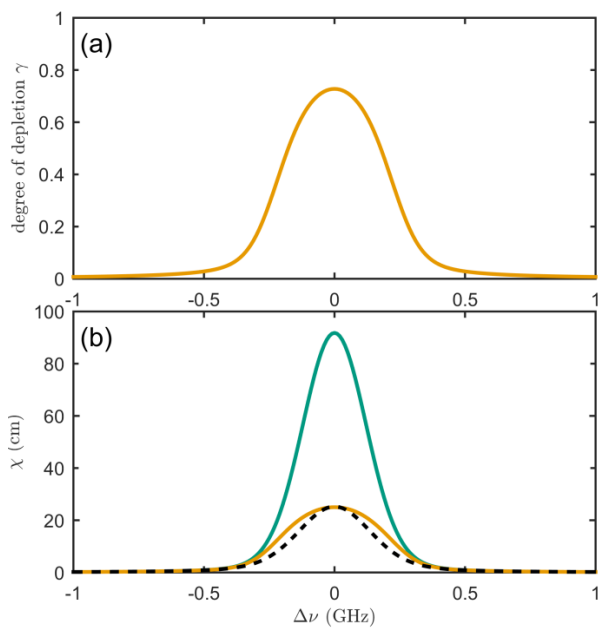

Fig. 9. Panel (a): the assessed degree of depletion as function of detuning for the pilot transition addressed at a pressure of 10 Torr and a power of 3 W. Panel (b): green curve: the area normalized Voigt absorption profile $\chi(\Delta \nu)$ for the same set of conditions; orange curve: the corresponding profile when the population in the vibrational ground state is depleted, calculated as $\chi(\Delta \nu) \cdot [1 - \gamma(\Delta \nu)]$; black dashed curve: a Voigt profile in the presence of optical saturation (with a degree of saturation of 7.5).

## 5. SUMMARY AND CONCLUSION

When spectroscopic absorption techniques are applied to molecular systems that experience a small probability for collisional induced deexcitation of vibrational excitations, the molecular system can be affected by depletion of its lowest vibrational state (a substantial fraction of the molecules can be transferred to excited vibrational states whereby the lowest vibrational level will be under-populated). When this takes place, an assessment of the concentration of the molecule under study based on a conventional model of absorption spectrometry will provide incorrect values.

As has been shown in an accompanying work [2], one system in which depletion can take place is methane in various buffer gases, in particular neat $N_2$. This restricts severely the ability to properly assess the concentration of this environmentally important species (and its various isotopologues) under various conditions.

To remedy this, a model for MAS that can incorporate the effect of depletion has been developed in this work. The formulation of the model has had two main aims; firstly, it should be as simple as possible, so it can be easily be used under a set of general conditions, and secondly it should provide results that can be verified by experiments, not least with the ones on methane that recently were made in our laboratory and that are presented in our accompanying paper [2]. Therefore the work includes a detailed discussion of all assumptions necessary to achieve a simple yet quantitatively good model considered the population of the vibrational states of methane excited by a narrow linewidth laser on a rovibrational transition (at around 3.4 μm) to a state in the pentad in the 1 – 100 Torr pressure range. However, the discussion can serve as a basis to adopt the model for different conditions and molecules.

The scrutiny of the energy level structure includes therefore features of methane, e.g. that its vibrational levels appear in so-called polyads and that it has three different spin isomers. But also features that are expected to be similar in most other molecules are considered, e.g. that for the range of pressures considered, there is a fast collision redistribution among various rotational levels and velocity groups within a given vibrational state. Based on these features several assumptions were made which implied that the methane molecules could be described by a simple three level system where the various levels represent the vibrational ground state, the dyad, and the pentad of the particular spin isomer being addressed. Although the assumptions imply inaccuracies of a few percent, it was considered as an advantage of describing a molecular system as a simple three-level system since it is not only simpler to use, but also significantly less details about the system are needed, and the various phenomena that take place can more easily be understood.

It was also shown in this work that this is the range in which excited molecules are simultaneously affected by two processes, which are both strongly pressure dependent. The first one is vibrational deexcitation through collisions with buffer gas molecules and the second one is drift (diffusion) in and out of the beam. Any quantitative model of the molecular system has therefore to include both these processes. To address the first, it could be concluded that methane is an exceptionally well-studied molecule for which a multitude of collisional rates constants have been assessed that can be used as input parameters of the model. To deal with the second, i.e. the transport of molecules in and out of the beam, the model incorporated diffusion according to Fick's law [30]. To correctly account for the depletion phenomenon, a diffusion term was added to the rate equation of each individual level.

To assess the influence of depletion on a measured absorption signal, an entity referred to as the assessed degree of depletion was introduced, representing the fraction by which the non-depleted absorption rate (or signal) is reduced by depletion. This can be used to not only assess how much a given absorption signal is decreased due to the depletion phenomena, but also to explore under which conditions conventional models of absorption are invalid. Also this parameter can be directly compared to experimental data, in particular the one presented in our accompanying paper [2].

The model was applied to methane molecules in trace concentrations in either neat $N_2$ or dry air (the latter consisting of 20% $O_2$ in $N_2$). Simulations were first performed of the population of the energy levels in the cavity in cylindrical coordinates (i.e. as a function of the radius from the axis of the laser field). The populations of molecules in the two laser addressed vibrational level were then used to calculate the amount of absorption of the laser light under various conditions.

The simulations showed that both collisional deexcitation and diffusion are of importance in the pressure range addressed; it was concluded that diffusion is the dominating deexcitation mechanism at low pressures, predominantly below around 10 Torr, while collisional deexcitation plays a progressively more important role for higher pressures. This gives the system a rather complex dependence on various entities.

For example, simulations of the spatial distribution of excited methane molecules indicated that at lower pressures, molecules can be in an excited vibrational state at significant (radial) distances from the laser beam. This implies that not all excited molecules that drift out of the laser-illuminated (LI) region, which can be deexcited either by collisions with buffer gas molecules or the cavity wall, will be replaced by molecules that are in a thermal equilibrium. Since the fraction of the molecules that drift into the LI region and are in an excited vibrational state depends on a number of system entities, this indicates that it is, in general, not appropriate to describe the exchange of molecules between the LI and the non-LI (NLI) region as a simple decay rate, as is done in many of the descriptions of laser-induced excitation of molecules that treat systems at sub-Torr pressures in which the exchange of molecules with the part of the sample that is outside the LI region often is modelled as a simple decay rate [13, 17].

It was also shown that the assessed degree of depletion increases with the excitation rate and thereby with laser power; the dependences are linear up to assessed degrees of depletion of around 20%, after which the dependence is weaker. It was predicted that the assessed degree of depletion can be significant, above 50% in neat $N_2$, for laser powers of 500 mW or above, for a range of pressures.

Moreover, the model also predicts that, for the methane system considered (methane in trace concentrations in neat $N_2$ or dry air), the maximum assessed degree of depletion is reached in the 10 - 30 Torr range (depending on the conditions considered). For low pressures (up to 5 Torr) the geometry of the cavity cell (primarily its diameter) plays a role for the deexcitation of methane molecules, and thereby the absorption signal. In these cases, the assessed degree of depletion can be reduced by decreasing the diameter of the sample cell. For higher pressures, however, (above 15 Torr for the pertinent situation) the molecules in the outer part of the cell are in thermal equilibrium. In these cases, and in particular for pressures above 50 Torr, the geometry of the cavity cell has no influence on the populations in the LI region and absorption process and the assessed degree of depletion has only a weak dependence on pressure. This pressure range should therefore be chosen if the non-linear pressure dependence of the depletion needs to be minimized. In these cases, it is also possible that the depletion phenomenon can be taken it into account by a simple linear pressure correction.

As the collisional deexcitation rates of methane are higher for collisions with $O_2$ than $N_2$, it is also predicted that there is more depletion of $CH_4$ in $N_2$ than in air, in particular at pressures above 50 Torr. This implies that systematic errors can appear if calibration is made in a cell that has a different buffer gas (e.g. methane in neat $N_2$) than the sample gas (e.g. methane in dry air). All this implies that absorption spectrometry of methane can demonstrate significant non-linear dependencies on pressure. It also underlines the importance of using a reference gas with similar type of buffer gas as the gas that is to be analyzed.

The model suggests that the assessed degree of depletion cannot only be reduced by decreasing the laser power, but also by adding buffer gas in which methane has a high collisional deexcitation rate (e.g. $O_2$ or $CO_2$), or, when low pressures are used, by using a sample cell with a smaller diameter.

The validity of the model is verified by experiments in methane, described in an accompanying work, performed at three different transitions, addressing molecules with all possible nuclear spin values, i.e. 0, 1, and 2, of two isotopologues ($^{12}CH_4$ and $^{13}CH_4$), in two types of buffer gases, neat $N_2$ and dry air, for a variety of pressures [2]. It was found that the agreement between the measured and theoretically predicted degrees of depletion was good and that all major features of the signals could be reproduced adequately, both qualitatively and quantitatively. This proves that although advanced multi-level descriptions can be used to model AS in the presence of depletion, it suffices, for the conditions considered, i.e. for pressures in the 1 – 100 Torr range, to model the system as a simple three-level system to which the transport of molecules in the system is coupled as diffusion according to Fick's law, applied to each level individually. This simplifies the use of the model when other molecules or other transitions in methane are addressed.

It is worth to note that for pressures lower than those addressed in this work (i.e. sub-Torr pressures), for which the model given here is not directly applicable, methane molecules will still be affected by collisional processes that equilibrate some types of levels in the various vibrational states, but not all. In such cases, since not all redistribution processes can be assumed to be sufficiently fast, the molecular system has to be modeled more meticulously, presumably by a model encompassing more levels.

Depletion effects are primarily assumed to affect systems in which the molecules are exposed to a high excitation rate for a continuous amount of time, as for example is the case in techniques such as DCEAS and NICE-OHMS [2], but less in techniques that excites the molecules temporarily, as for example is the case in cavity ring-down spectroscopy (CRDS), since discontinued interactions will reduce the amount of depletion in the system. However, it is feasible that CRDS is affected by the depletion phenomenon when methane is addressed, in particular if a high duty cycle is used, as for example is the case in the frequency-agile rapid scanning, cavity ring-down spectroscopy technique [49].

**Funding Information.** Swedish research Council (VR) (2015-04374). The Kempe foundations (JCK 1317.1).

**Acknowledgment**. We thank Michael Bradley for assistance with the diffusion equation and Umeå University's program "Strong research environments" for support

## References

[1] Ye J, Lynn TW. Applications of optical cavities in modern atomic, molecular, and optical physics. Adv Atom Mol Opt Phy. 2003;49:1-83.
[2] Hausmaninger T, Zhao G, Ma W, Axner O. Depletion of the vibrational ground state of $CH_4$ in absorption spectroscopy at 3.4 µm in $N_2$ and air in the 1 to 100 Torr range. J Quant Spectrosc Radiat Tr. 2017;205:59-70.
[3] Borde CJ, Hall JL, Kunasz CV, Hummer DG. Saturated absorption-line shape - Calculation of transit-time broadening by a perturbation approach. Phys Rev A. 1976;14:236-63.
[4] Sargent III M, O. Scully M, Lamb Jr. WE. Laser Physics: Addison-Wesley; 1977.
[5] Loudon R. The Quantum Theory of Light. 3 ed: Oxford U. Press; 2000.
[6] Ma WG, Foltynowicz A, Axner O. Theoretical description of Doppler-broadened noise-immune cavity-enhanced optical heterodyne molecular spectroscopy under optically saturated conditions. J Opt Soc Am B. 2008;25:1144-55.
[7] Demtröder W. Laser Spectroscopy. 2nd ed. Berlin: Springer Verlag; 1996.
[8] Romanini D, Dupre P, Jost R. Non-linear effects by continuous wave cavity ringdown spectroscopy in jet-cooled $NO_2$. Vib Spectrosc. 1999;19:93-106.
[9] Labazan I, Rudic S, Milosevic S. Nonlinear effects in pulsed cavity ringdown spectroscopy of lithium vapour. Chem Phys Lett. 2000;320:613-22.
[10] Foltynowicz A, Ma WG, Schmidt FM, Axner O. Doppler-broadened noise-immune cavity-enhanced optical heterodyne molecular spectrometry signals from optically saturated transitions under low pressure conditions. J Opt Soc Am B. 2008;25:1156-65.
[11] Milonni PW, Eberly JH. Laser Physics: John Wiley, New York, NY 2010.
[12] Sadiek I, Friedrichs G. Saturation dynamics and working limits of saturated absorption cavity ringdown spectroscopy. Physical Chemistry Chemical Physics. 2016;18:22978-89.
[13] Ye J. Ultrasensitive high resolution laser spectroscopy and its application to optical frequency standards [Ph.D. Thesis]. Boulder, Colorado: University of Colorado; 1997.
[14] Ma LS, Ye J, Dube P, Hall JL. Ultrasensitive frequency-modulation spectroscopy enhanced by a high-finesse optical cavity: theory and application to overtone transitions of $C_2H_2$ and $C_2HD$. J Opt Soc Am B. 1999;16:2255-68.
[15] Bucher CR, Lehmann KK, Plusquellic DF, Fraser GT. Doppler-free nonlinear absorption in ethylene by use of continuous-wave cavity ringdown spectroscopy. Appl Opt. 2000;39:3154-64.
[16] Lisak D, Hodges JT. High-resolution cavity ring-down spectroscopy measurements of blended H2O transitions. Appl Phys B. 2007;88:317-25.
[17] Axner O, Ma WG, Foltynowicz A. Sub-Doppler dispersion and noise-immune cavity-enhanced optical heterodyne molecular spectroscopy revised. J Opt Soc Am B. 2008;25:1166-77.
[18] Mills AA, Siller BM, McCall BJ. Precision cavity enhanced velocity modulation spectroscopy. Chem Phys Lett. 2010;501:1-5.
[19] Urban W, Lin JX, Subramaniam VV, Havenith M, Rich JW. Treanor pumping of CO initiated by CO laser excitation. Chem Phys. 1989;130:389-99.
[20] Yardley JT, Fertig MN, Moore CB. Vibrational deactivation in methane mixtures. J Chem Phys. 1970;52:1450-3.
[21] Boursier C, Menard J, Doyennette L, Menard-Bourcin F. Rovibrational relaxation of methane in $CH_4$-$N_2$ mixtures: Time-resolved IR-IR double-resonance measurements at 193 K and kinetic Modeling. J Phys Chem A. 2003;107:5280-90.
[22] Whiticar MJ. Carbon and hydrogen isotope systematics of bacterial formation and oxidation of methane. Chem Geol. 1999;161:291-314.
[23] Hakala JA. Use of stable isotopes to identify sources of methane in Appalachian Basin shallow groundwaters: a review. Environ Sci Process Impact. 2014;16:2080-6.
[24] Zhao G, Hausmaninger T, Ma WG, Axner O. Whispering-gallery-mode laser-based noise-immune cavity-enhanced optical heterodyne molecular spectrometry. Opt Lett. 2017;42:3109-12.
[25] Klaassen JJ, Coy SL, Steinfeld JI, Abel B. Direct measurement of rotational and vibrational-relaxation in methane overtone levels by time-resolved infrared double-resonance spectroscopy. J Chem Phys. 1994;101:10533-47.
[26] Strekalov ML. Rotational energy relaxation in $CH_4$ and $CH_4$-He, Ar collisions calculated from coherent and stimulated Raman spectroscopy data. Mol Phys. 2002;100:1049-56.
[27] Aliat A, Chikhaoui A, Kustova EV. Nonequilibrium kinetics of a radiative CO flow behind a shock wave. Phys Rev E. 2003;68.
[28] Menard-Bourcin F, Boursier C, Doyennette L, Menard J. Rotational and vibrational relaxation of methane excited to $2v_3$ in $CH_4/H_2$ and $CH_4$/He mixtures at 296 and 193 K from double-resonance measurements. J Phys Chem A. 2005;109:3111-9.
[29] Menard-Bourcin F, Doyennette L, Menard J, Boursier C. Time-resolved IR-IR double resonance measurements in methane excited to $2v_3(F_2)$. J Phys Chem A. 2000;104:5444-50.
[30] Bird RB, Stewart WE, Lightfoot EN. Transport Phenomena. John Wiley & Sons.; 1976.
[31] Schilt S, Besson JP, Thevenaz L. Near-infrared laser photoacoustic detection of methane: the impact of molecular relaxation. Appl Phys B. 2006;82:319-28.
[32] Doyennette L, Menard-Bourcin F, Menard J, Boursier C, Camy-Peyret C. Vibrational energy transfer in methane excited

to 2ν$_3$ in CH$_4$-N$_2$/O$_2$ mixtures from laser-induced fluorescence measurements. J Phys Chem A. 1998;102:3849-55.
[33] Brown LR, Sung K, Benner DC, Devi VM, Boudon V, Gabard T, et al. Methane line parameters in the HITRAN2012 database. J Quant Spectrosc Radiat Tr. 2013;130:201-19.
[34] Rothman LS, Gordon IE, Babikov Y, Barbe A, Benner DC, Bernath PF, et al. The HITRAN2012 molecular spectroscopic database. J Quant Spectrosc Radiat Tr. 2013;130:4-50.
[35] Brown LR, Benner DC, Champion JP, Devi VM, Fejard L, Gamache RR, et al. Methane line parameters in HITRAN. J Quant Spectrosc Radiat Tr. 2003;82:219-38.
[36] Hepp M, Winnewisser G, Yamada KMT. Conservation of the nuclear-spin modification of CH$_4$ in the cooling process by supersonic jet expansion. J Mol Spectros. 1994;164:311-4.
[37] Yardley JT, Moore CB. Intramolecular vibration-to-vibration energy transfer in carbon dioxide. J Chem Phys. 1967;46:4491-5.
[38] Yardley JT, Moore CB. Vibration - vibration and vibration - translation energy transfer in methane-oxygen mixtures. J Chem Phys. 1968;48:14-7.
[39] Yardley JT, Moore CB. Vibrational energy transfer in methane. J Chem Phys. 1968;49:1111-25.
[40] Menard-Bourcin F, Boursier C, Doyennette L, Menard J. Rovibrational energy transfer in methane excited to 2ν$_3$ in CH$_4$-N$_2$ mixtures from double-resonance measurements. J Phys Chem A. 2001;105:11446-54.
[41] Boursier C, Menard J, Menard-Bourcin F. Vibrational relaxation of methane by oxygen collisions: Measurements of the near-resonant energy transfer between CH$_4$ and O$_2$ at low temperature. J Phys Chem A. 2007;111:7022-30.
[42] Menard-Bourcin F, Menard J, Boursier C. Temperature dependence of rotational relaxation of methane in the 2ν$_3$ vibrational state by self- and nitrogen-collisions and comparison with line broadening measurements. J Mol Spectros. 2007;242:55-63.
[43] Cacciani P, Cosleou J, Khelkhal M, Cermak P, Puzzarini C. Nuclear Spin Conversion in CH$_4$: A Multichannel Relaxation Mechanism. J Phys Chem A. 2016;120:173-82.
[44] Simeckova M, Jacquemart D, Rothman LS, Gamache RR, Goldman A. Einstein A-coefficients and statistical weights for molecular absorption transitions in the HITRAN database. J Quant Spectrosc Radiat Tr. 2006;98:130-55.
[45] Barreiro N, Peuriot A, Santiago G, Slezak V. Water-based enhancement of the resonant photoacoustic signal from methane-air samples excited at 3.3 μm. Appl Phys B. 2012;108:369-75.
[46] Chapovsky PL, Hermans LJF. Nuclear spin conversion in polyatomic molecules. Annu Rev Phys Chem. 1999;50:315-45.
[47] Pfeiffer Vacuum. The Vacuum Technology Book Volume II. Pfeiffer Vacuum GmbH2013.
[48] Marrero TR, Mason EA. Gaseous Diffusion Coefficients. Journal of Physical and Chemical Reference Data. 1972;1:3-118.
[49] Truong GW, Douglass KO, Maxwell SE, van Zee RD, Plusquellic DF, Hodges JT, et al. Frequency-agile, rapid scanning spectroscopy. Nat Photon. 2013;7:532-4.